\documentstyle[12pt]{article}
\def\ba{\begin{eqnarray}}
\def\ea{\end{eqnarray}}

\def\lb{\label}
\def\be{\begin{equation}}
\def\ee{\end{equation}}

\begin{document}

\title{Complete Calabi-Yau metrics from Kahler metrics in D=4}
\author{Mauricio Leston \thanks{Instituto de Astronomia y Fisica del Espacio (IAFE), Buenos Aires, Argentina; mauricio@iafe.uba.a} and  Osvaldo P. Santillan \thanks{Departamento de Matematica, FCEyN, Universidad de Buenos Aires, Buenos Aires, Argentina;
firenzecita@hotmail.com and osantil@dm.uba.ar} }
\date {}
\maketitle

\begin{abstract}
In the present work, a family of Calabi-Yau manifolds with a local
Hamiltonian Killing vector is described in terms of a non linear equation
whose solutions determine the local form of the geometries. The main assumptions
are that the complex $(3,0)$-form is of the form $e^{ik}\overline{\Psi}$,
where $\overline{\Psi}$ is preserved by the Killing vector, and that the
space of the orbits of the Killing vector is, for fixed value
of the momentum map coordinate, a complex 4-manifold, in such a way that the
complex structure of the 4-manifold is part of the complex structure of the
complex 3-fold. The family
considered here include the ones considered in \cite{Fayyazuddin}-\cite{Chinos}
as a particular case. We also present an explicit example with holonomy
exactly SU(3) by use of the linearization introduced in \cite{Fayyazuddin},
which was considered in the context of D6 branes wrapping a complex 1-cycle in a hyperkahler 2-fold.

\end{abstract}

\tableofcontents

\section{Introduction}

    The development of the subject of Calabi-Yau (CY) manifolds is an illustrative example
of the interplay between algebraic geometry and string theory.
On the one hand, CY spaces are interpreted as internal spaces of string and M-theory
giving supersymmetric field theories after compactification.
In fact, CY 3-folds may provide compactifications which are more realistic than the
ones corresponding to other Ricci-flat manifolds such as $G_2$ holonomy spaces, for
which the generation of chiral matter and non abelian gauge symmetries seems harder
(but not impossible) to achieve. On the other hand, string theory compactifications
stimulated several new trends in the algebro-geometrical aspects of CY spaces,
an example is the subject of mirror symmetry.

    By definition a CY manifold is a compact Kahler n-dimensional manifold with
vanishing first Chern class. The Yau proof of the Calabi conjecture implies that
these manifolds admit a Ricci-flat metric and their holonomy is reduced from SO(2n) to
SU(n) \cite{Yau}.  Although compact Ricci-flat metrics exist, no explicit expressions
have been found. The main technical problem for that is that a compact Ricci-flat
metric does not admit globally defined Killing vectors (leaving aside the possibility
to have trivial flat $U(1)$ factors), and the absence of continuous symmetries makes
the task of solving the Einstein equations explicitly really hard. For the non compact
case, the definition usually adopted is that
a CY manifold is a Ricci-flat Kahler manifold, which also implies that the holonomy is
reduced to SU(n) or to a smaller subgroup.
In this case several Calabi-Yau metrics with isometries have been found in
\cite{Candelas}- \cite{Yau2}, and \cite{Martelli1}-\cite{Edelstein}. Some of
these metrics posses conical singularities but in some cases these singularities
have been resolved to give complete metrics.

 Although non compact Calabi-Yau metrics are not suitable for studying
compactification in string theory, they have several applications in
mathematical and theoretical physics. For instance, the localization
techniques pioneered by Kontsevich \cite{kontzo}-\cite{chun}  to calculate
Gromov-Witten invariants is more easy to implement in the non compact case
and sometimes these invariants may been calculated for arbitrary genus. Also,
it was conjectured in \cite{gopvaf} that Chern-Simmons on $S^3$ is equivalent
to topological strings on the resolved conifold $T^* S^3$, which is Calabi-Yau.
These has been generalized in \cite{Aganagic} where it is shown that for some three
dimensional manifold $M$ the space $T^* M$, is Calabi-Yau and it was conjectured
that Chern-Simmons on $M$ is dual to topological strings propagating in $T^*M$ (See
\cite{Marino} for a nice review).

    In view of the above discussion, to find general methods
for constructing non compact CY metrics with isometries is a task of interest.
An step in that direction was initiated by Fayyazuddin in \cite{Fayyazuddin} where
the supergravity backgrounds corresponding to D6 branes wrapping a complex submanifold
inside a 4-dimensional hyperkahler space were characterized in terms of a single linear
equation. It was also shown in that reference that the uplift to eleven dimensions results in a purely
geometrical background of the form $M_{1,4} \times Y_6$ where $Y_6$ is a Calabi-Yau
space. The Ricci-flat Kahler metric on $Y_6$ is therefore determined by this linear
equation, which is expressed in term of the laplacian over the curved hyperkahler
space the branes wrap. For all these geometries there is a U(1) isometry preserving
the whole $SU(3)$ structure (which is in particular hamiltonian and therefore it
defines a momentum map local coordinate) such that space formed by the orbits of
the Killing vector is, for fixed values of the momentum map coordinate, a Kahler manifold.
The Fayyazuddin construction was reconsidered in \cite{osvaldo} where it was shown that the assumption
that the quantities defining the geometry vary over a complex submanifold may be relaxed
without violating the Calabi-Yau condition. The resulting geometries were described in
terms of a non linear equation, which reduce to the Fayyazuddin one if the quantities
describing the geometry vary over a complex submanifold. The non linear operator is
defined in terms of the metric of the hyperkahler space, in fact, this method can be
interpreted as a solution generating technique which starts with a hyperkahler metric
and gives a non compact Calabi-Yau metric as outcome.

    The two approaches mentioned above have been used to find non-trivial Calabi-Yau metrics with holonomy exactly SU(3).
Nevertheless, none of these examples were complete metrics. This situation
was substantially improved in \cite{Chinos} where isometries which do not
preserve the SU(3) structure, but just the metric $g_6$ and the Kahler two form
$\omega_6$, were considered. These authors showed that one may start with a hyperkahler
structure as well and construct complete Calabi-Yau metrics. In particular, the
resolution of the $Y^{p,q}$ cone found in \cite{sun}-\cite{yi} was rediscovered in
these terms. The calculations made in \cite{Chinos} are impressive, but there is a
striking fact there that motivates
the present note, which is the following. The best results obtained in
\cite{Chinos} are obtained in terms of the flat hyperkahler structure on $R^4$, in particular,
the resolution of the Ricci-flat cone over $Y^{p,q}$. Instead, for a curved
hyperkahler structure, the resulting equations seem harder to solve and more
restricted solutions are found, or even no solutions at all. One may wonder
if a method for constructing Calabi-Yau metrics without the use of initial
hyperkahler structures may be developed, which may allow us to avoid this kind of problems.
In the present such a method will be presented and family of Calabi-Yau geometries
characterized by a single non linear equation which is not necessarily related to a
hyperkahler metric. It should be emphasized that there is \emph{nothing} wrong with
the use of hyperkahler structures as initial input. What the present letter shows is
that this is just optional.

The organization of the present work is as follows.
In section 2.1 generalities about SU(3) structures are reviewed.
In section 2.2  the SU(3) structures with a Hamiltonian Killing vector,
that is, a Killing vector preserving also the Kahler form are characterized. In section 2.3 a family of Calabi-Yau metrics of this type
is presented, for which the complex
$(3,0)$ form is of the form $\Psi=e^{ik}\overline{\Psi}$ in such a way that
$\overline{\Psi}$ is preserved by the Killing vector but $\Psi$ may not be
preserved due to the phase factor. In section 3.1 and 3.2 it is explained that the metrics
considered in \cite{osvaldo} and \cite{Fayyazuddin} belong to the family of section 2.3. In section
3.3 an example where the Fayyazuddin linearization \cite{Fayyazuddin} works properly
is worked out explicitly and a non-trivial Calabi-Yau metric is obtained as outcome.
In section 3.4 we also show that the results of \cite{Chinos} belong to the family
constructed here. Section 4 contains the discussion of the results obtained.

\section{Calabi-Yau metrics with Hamiltonian isometries}

\subsection{The general form of the $SU(3)$ structure}

    In this subsection a large family of Calabi-Yau (CY) manifolds in dimension 6 with an isometry group
with orbits of codimension one will be characterized. It will be assumed that the Killing vector
$V$ corresponding to this isometry preserve not only the metric, but the full Kahler
two form $\omega_6$. It will be convenient to give an operative definition of CY manifolds
in six dimensions first, for more details see for instance \cite{Besse}. Roughly speaking, a Calabi-Yau manifold $M_6$ is Kahler manifold,
thus complex sympletic, which in addition admits a Ricci-flat metric $g_6$. This definition
means that there exist an endomorphism of the tangent space $J: TM_6\to TM_6$
such that $J^2=-I_d$ and for which $g_6(X, J Y)=-g_6(J X, Y)$ being $X$ and $Y$  arbitrary
vector fields. It is commonly said that the metric $g_6$ is hermitian with respect to $J$
and the tensor $(g_6)_{\mu\alpha}J^{\alpha}_{\nu}$ is skew symmetric, therefore locally it
defines a 2-form
\be\lb{r}
\omega_6=\frac{1}{2}\;(g_6)_{\mu\alpha}J^{\alpha}_{\nu} dx^{\mu}\wedge dx^{\nu}.
\ee
Here $x^{\mu}$ is a local choice of coordinates for $M_6$. The endomorphism $J$ it is called
an almost complex structure. If the Nijenhuis tensor
$$
N(X, Y)=[X, Y]+J\;[X, J \;Y]+J\;[J\;X, Y]-[J\;X, J\;Y],
$$
vanishes identically then the tensor $J$ will be called a complex
structure and $M_6$ a complex manifold. This is the case for any CY manifold. The
Newlander-Niremberg theorem states that there is an atlas of charts for $M_6$ which
are open subsets in $C^{n}$, in such a way  that the transition maps are holomorphic
functions. These local charts are parameterized by complex coordinates
$(z_i, \overline{z}_i)$ with $i=1,2,3$ for which the complex structure looks like
\be\lb{lula}
J_{i}^{j}=-J_{\overline{i}}^{\overline{j}}=i\delta_i^j,\qquad J_{i}^{\overline{j}}= J^{i}_{\overline{j}}=0,
\ee
and for which the metric and the 2-form (\ref{r}) are expressed as follows
\be\lb{expro}
g_6=(g_6)_{i\overline{j}} \;dz_{i} \otimes d\overline{z}_{j},
\ee
\be\lb{expro2}
\omega_6=\frac{i}{2}\;(g_6)_{i\overline{j}} \;dz_{i}\wedge d\overline{z}_{j}.
\ee
The form (\ref{expro2}) is called of type $(1,1)$ with respect to $J$, while a generic
2-form containing only terms of the form $(dz_i\wedge dz_j)$ or
$(d\overline{z}_i\wedge d\overline{z}_j)$ will be called of type $(2, 0)$ or $(0, 2)$,
respectively.  In addition a Calabi-Yau manifold is sympletic with respect to $\omega_6$,
in other words $d\omega_6=0$. A complex manifold which is sympletic with respect to (\ref{r})
is known as a Kahler manifold, thus CY spaces  are all Kahler. The Kahler condition itself
implies that the holomy is reduced from $SO(6)$ to $U(3)$. Furthermore, the fact that $g_6$
is Ricci-flat is equivalent to the existence of a 3-form
\be\lb{s}
\Psi=\psi_{+}+i \;\psi_{-},
\ee
of type $(3,0)$ with respect to $J$, satisfying the compatibility conditions
\cite{Chiozzi}
\be\lb{condi}
\omega_6\wedge \psi_{\pm}=0,\qquad \psi_{+}\wedge \psi_{-}=\frac{2}{3}\omega_6\wedge \omega_6\wedge \omega_6\simeq dV(g_6),
\ee
and which is closed, i.e,
\be\lb{cob}
d\psi_{+}=d\psi_{-}=0.
\ee
The relations (\ref{condi}) can be expressed in more compact way as
\be\lb{condil}
\omega_6\wedge \Psi=0,\qquad \Psi\wedge \overline{\Psi}=\frac{1}{3}\omega_6\wedge \omega_6\wedge \omega_6\simeq dV(g_6).
\ee
In the formula (\ref{condil}) $dV(g_6)$ denote the volume form of $g_6$. In the situations
described in (\ref{cob})  the holonomy is further reduced from $U(3)$ to $SU(3)$, thus CY
 manifolds are of $SU(3)$ holonomy. The converse of these statements are also true, that is,
  for any Ricci-flat Kahler metric in D=6 there will exist an SU(3) structure
  $(\omega_6, \Psi)$ satisfying (\ref{condil}) and also
\be\lb{co}
d\omega_6=d\Psi=0.
\ee
The knowledge SU(3) structure determine univocally metric $g_6$. In fact, the task to find
complex coordinates for a given CY manifold may be not simple, but there always exists
a tetrad basis $e^a$ with $a=1,..,6$ for which the SU(3) structure is expressed as
\begin{equation}
\omega_6=\frac{i}{2}(E_1\wedge{\overline{E}}_1 +E_2\wedge{\overline{E}}_2+E_3\wedge{\overline{E}}_3,\label{estruct2})
\end{equation}
\begin{equation}
\Psi=E_1\wedge{E}_2\wedge{E}_3,\label{estruct3}
\end{equation}
where $E_i\equiv{e_j}+i\;e_{j+1}$ ($j=1,3,5$), and for which the metric is
\be\lb{metfv}
g_6=E_1\otimes{\overline{E}}_1 +E_2\otimes{\overline{E}}_2+E_3\otimes{\overline{E}}_3
\ee
 Note that the multiplication by a phase factor $E_i\to e^{i\; k} E_i$ does not change
the metric and induce the transformation
$\Psi \to e^{3\;i\;k} \Psi$ on the (3,0) form.
This phase transformation does not alter the conditions (\ref{condil}),
this fact will be important in the following.

\subsection{Kahler structures with Hamiltonian
 isometries}

      The description given above just collects general facts about CY manifolds. In the
following we will assume that our CY manifold $M_6$ is equipped with a metric $g_6$ in
such a way that there is a Killing vector $V$ preserving $g_6$ and the Kahler form
$\omega_6$. In this situation there exists a local coordinate system $(\alpha, x^i)$
with $i=1,..,5$ for which $V=\partial_{\alpha}$ and for which the metric tensor $g_6$
takes the following form
\be\lb{CY}
g_6=\frac{(d\alpha+A)^2}{H^2}+H g_5,
\ee
where the function $H$, the one form $A$ and the metric tensor $g_5$ are independent on the
coordinate $\alpha$. Thus these objects live in a 5-dimensional space
which we denote $M_5$.  The metric $g_5$ appearing in (\ref{CY}) can be expressed as
$g_5=e^a\otimes e^a$ with $a=1,..,5$ for some basis of $\alpha$-independent 1-forms $e^a$.
Then, if $V$ also preserves the Kahler form $\omega_6$ (as we are assuming), one has the
decomposition
\be\lb{su3}
\omega_6=\omega_4+\frac{1}{\sqrt{H}}e^5\wedge (d\alpha+A).
\ee
Here the 1-form $e^5$ is by definition
\be\lb{analog1}
\frac{e^5}{\sqrt{H}}=i_{\partial_{\alpha}}\omega_{6},
\ee
$i_{V}$ denoting the contraction with the vector field $V$. The elementary formula in
differential geometry
\be\lb{analog2}
d_5(i_{\partial_{\alpha}}\omega_{6})=\pounds_{\partial_{\alpha}}\omega_6-i_{\partial_{\alpha}}\;d\omega_6,
\ee
together with (\ref{analog1}) implies that
\be\lb{analog3}
d_5(\frac{e^{5}}{\sqrt{H}})=\pounds_{\partial_{\alpha}}\omega_6-i_{\partial_{\alpha}}\;d\omega_6.
\ee
Here $d_5=\partial_i\; dx^i$ and $\pounds_{\partial_{\alpha}}$ is the Lie derivate along
the vector $\partial_{\alpha}$. But the vector $\partial_{\alpha}$, by assumption,
preserves $\omega_6$ and $\omega_6$ is closed, thus the right hand side of (\ref{analog3})
vanishes and
\be\lb{s2}
d_5(\frac{e^{5}}{\sqrt{H}})=0.
\ee
The last relation can be integrated, at least locally, to obtain that
\be\lb{retu}
e^5=\sqrt{H}\;dy,
\ee
$y$ being some function of the coordinates $x^i$ parameterizing $M_5$, which is known as the momentum
map of the isometry.  At least locally, one can take the function $y$ defined in (\ref{retu})
as one of the coordinates, which leads to the decomposition $M_5=M_4\times R_{y}$ and
$d_5=d_4+\partial_y \;dy$. The metric (\ref{CY}) in this coordinates becomes
\be\lb{CY2}
g_6=\frac{(d\alpha+A)^2}{H^2}+H^2 dy^2+H\;g_4(y),
\ee
where the tensor $H\;g_4(y)$ will be determined below under certain additional assumptions.
The Kahler form is
\be\lb{sisv}
\omega_6=\omega_4(y)+dy\wedge (d\alpha+A).
\ee
The next task will be to find specific examples of this type of structures.

\subsection{Calabi-Yau metrics with Hamiltonian isometries}

    In this subsection, the generic Kahler structure described above
will be extended to an specific family of Calabi-Yau structures. The main assumption will
be that, for fixed values of the coordinates ($\alpha$, $y$), the resulting 4-manifold is
complex, and that the two form $\omega_4$ appearing in (\ref{sisv}) is of type $(1,1)$
with respect a complex coordinate system for this manifold. This may be paraphrased by saying that the complex structure
of the complex 4-manifold is part of the complex structure of the Ricci-flat Kahler 6-manifold.
By denoting the complex coordinates as
$(z_1,z_2,\overline{z}_1,\overline{z}_2)$, the main assumption implies that (\ref{CY2}) may be expressed as
\be\lb{CY34}
g_6=\frac{(d\alpha+A)^2}{H^2}+H^2 dy^2+ H\; g_4(y)_{z_i\overline{z}_j}dz_i\otimes{d}\overline{z}_j,
\end{equation}
and the dependence on the coordinate $y$ is only as a parameter.

   In order to extend the Kahler structure given above to an $SU(3)$ structure,
an anzatz for the form $\Psi$ of (\ref{s}) is needed, in such a way
that the compatibility conditions (\ref{condil})
are identically satisfied. By analogy with the choice \cite{Chinos} we propose
the following form for $\Psi$
\begin{equation}
\Psi=e^{iK}\Omega_4\;\wedge[H\;dy + i\frac{(d\alpha+A)}{H}],\label{anzat2}
\end{equation}
$K$ being a function that may depend $\alpha$ and varying over $M_5$. The remaining quantities appearing
in \ref{anzat2} are assumed to be $\alpha$-independent. The compatibility
conditions (\ref{condil}) are then satisfied if and only if
\begin{equation}
2\;\omega_4\wedge\omega_4 =\Omega_4\wedge\overline{\Omega}_4=4\;\det(H\;g_4)dz_1\wedge dz_2\wedge d\overline{z}_1\wedge d\overline{z}_2,
\label{relal}
\end{equation}
This relation is, for fixed value of the coordinate $y$, the same as the compatibility
condition for SU(2) structures. It is a standard fact that if there is complex
coordinate system
for which $\omega_4$ is of type (1,1), then  $\Omega_4$ is of type (2,0) with respect to it.
This means that
$$
\Omega_4=H\;f\;dz_1\wedge dz_2,
$$
$f$ being a function independent on $\alpha$ and varying over $M_5$ and the factor $H$ in front
is just by convenience. The compatibility condition (\ref{relal}) implies that
\begin{equation}2\;\omega_4\wedge\omega_4 =\Omega_4\wedge\overline{\Omega}_4
=H^2\;f^2\;dz_1\wedge{d}\overline{z}_1\wedge dz_2\wedge{d}\overline{z}_2,
\label{rela}
\end{equation}
and by comparing (\ref{relal}) with (\ref{rela}) one obtains
\begin{equation}
H^2\;f^2=4\;\det(H\;g_4).\label{relaHg}
\end{equation}
Taking into account all these relations and (\ref{anzat2}) it follows easily that
\begin{equation}\label{omga}
\Psi=e^{i\;K}H^2\;f\;dz_1\wedge{d}z_2\wedge[\;dy + \frac{i(d\alpha + A)}{H^2}\;].
\end{equation}
The next task is to fix the unknown quantities $A$, $H$, $f$ and $K$ by the Calabi-Yau
condition (\ref{co}). The first one applied to (\ref{sisv}) gives
\begin{equation}
d_4\omega_4(y)=0,\label{kahlericidad}
\end{equation}
and
\begin{equation}
d_4A=\partial_y\omega_4,\label{potencial}
\end{equation}
Note that the equation (\ref{kahlericidad}) imply, for fixed value of $y$, that $H\; g_4$
is not only complex but also \emph{Kahler}. The second (\ref{co}) gives several equations,
corresponding to the vanishing of each component of $d\Psi$. The vanishing of the terms with
$(dz_1\wedge dz_2\wedge dy\wedge{d}\alpha)$ imply that
\begin{eqnarray}
K_y=0,\\
H^2\;f\;\partial_{\alpha}K-f_y=0.
\end{eqnarray}
The second equation implies that $K=K_0+\alpha{K}_1$, with $K_0$ and $K_1$ independent of $y$.
By combining this with the first one it is obtained that
\begin{equation}
H^2\;f\;K_1=f_y. \label{cort}
\end{equation}
The terms of the form $(dz_1\wedge dz_2\wedge d\alpha\wedge d\overline{z}_i)$  vanish if and
only if
\begin{eqnarray}
\overline{\partial}K_1=0,\\
-f\;\overline{\partial}K_0+i\;\overline{\partial}f + f\;{K}_1\;\overline{A}=0.
\end{eqnarray}
Since $K_1$ is real and $y$-independent, the first of these equations imply that it is a
constant, which can be set to $0$, $1$ without loosing generality. The case $K_1=0$
correspond to a Killing vector
preserving the whole SU(3) structure, which is the case considered in \cite{osvaldo}.
But for the moment we focus in the case $K=1$. In this case the last equation implies that
\begin{equation}
d_4^c f=f\;d_4K_0-K_1\;f\;A.\label{ecppal}
\end{equation}
For these cases the terms with
$(dz_1\wedge dz_2\wedge dy\wedge d\overline{z}_i)$ vanish when
\begin{equation}
d_4^c f_y=-K_1\;\partial_y(f\;A).\label{ecppal2}
\end{equation}
An immediate consequence the last two equation is
\begin{equation}
d_4K_0=0.
\end{equation}
Inserting this relation into (\ref{ecppal}) gives
\begin{equation}
d^c_4(\log f)=-A.
\end{equation}
By taking $d_4$ in both sides of the last equation and using (\ref{potencial})
it is seen that
\begin{equation}
d_4d_4^c (\log f)=-\partial_y\omega_4.\label{potencial2}
\end{equation}
But the condition (\ref{kahlericidad}) implies that the complex 4-dimensional
manifold $M_4$ is also a \emph{Kahler} manifold,
with $\omega_4$ being the Kahler form. Therefore $\omega_4$ has a Kahler
potential $G$, that is, $\omega_4=d_4d_4^cG$. The equation (\ref{potencial2}) imply that
\begin{equation}
f=U(z_1, z_2)\;e^{-G_y},\label{exp}
\end{equation}
with $U(z_1,z_2)$ an arbitrary holomorphic function. In addition, equation (\ref{cort}) gives that $H^2=G_{yy}$, and by combining this with
(\ref{rela}) and (\ref{exp}) it is obtained that
\begin{equation}\label{kitail}
U(z_1,z_2)(e^{-2\;G_y})_y=32\;(G_{1\overline{1}}G_{2\overline{2}}-G_{1\overline{2}}G_{2\overline{1}}),
\end{equation}
and that $H^2=G_{yy}$, with $G_{i\overline{j}}=\partial_i\partial_{\overline{j}}G$. But the holomorphic function can be absorbed by a holomorphic coordinate
change $z_i^`=f_i(z_1,z_2)$, thus there exist always a local coordinate system such that (\ref{kitail}) takes the form
\begin{equation}\label{kitai}
(e^{-2\;G_y})_y=32\;(G_{1\overline{1}}G_{2\overline{2}}-G_{1\overline{2}}G_{2\overline{1}}),
\end{equation}
In this way, all the quantities appearing in the six
dimensional metric are expressed in terms of $G$. Explicitly, the Calabi-Yau metric is
\begin{equation}\label{metricofin}
g_6=\frac{(d\alpha + d_4^cG_y)^2}{G_{yy}} + G_{yy}dy^2 + 2\;G_{i\overline{j}}\;dz_i\otimes{d}\overline{z}_j.
\end{equation}
For $K_1=0$, a calculation completely analogous to the one given above shows that the metric
is still (\ref{metricofin}) but in this case $G$ is given by
\begin{equation}\label{kitai2}
G_{yy}=8(G_{1\overline{1}}G_{2\overline{2}}-G_{1\overline{2}}G_{2\overline{1}}).
\end{equation}
Note that in both cases $K_1=0,1$ the metric is determined in terms of a single function $G$.

    It should be mentioned that the method described by (\ref{kitai}) or (\ref{kitai2}) may be
generalized to arbitrary complex dimensions in straightforward manner. The resulting metrics will be described by
(\ref{metricofin}) but the function $G$ will depend on n-complex coordinates $z_i$ with $i=1,..,n$ and will be the solution
of
 \begin{equation}\label{kitai9}
(e^{-2\;G_y})_y=2^{2n+1}\;\det(G_{i\overline{j}}),
\end{equation}
for $K_1=1$ and of
\begin{equation}\label{kitai29}
G_{yy}=2^{n+1}\;\det(G_{i\overline{j}}),
\end{equation}
for $K_1=0$, $\det(G_{i\overline{j}})$ being the determinant of the matrix whose entries are  the second derivatives of $G$ of type $(1,1)$. The resulting metric (\ref{metricofin}) will have $(n+1)$ complex dimensions but in the following we will keep considering the case
$n=2$.

\section{Solutions related to hyperkahler structures}

   In the following sections, the connection between the solution generating technique
given by (\ref{kitai})-(\ref{kitai2}) and the
known ones given in \cite{Fayyazuddin}-\cite{Chinos} will be detailed. The assumptions for obtaining the CY metrics
(\ref{kitai})-(\ref{kitai2}) were the following: there is an isometry
preserving the CY metric and the Kahler two form; the complex 3-form has the generic expression (\ref{anzat2})
; the manifold obtained for fixed values of $y$ and $\alpha$ is complex, in such a way that the metric is of the form
(\ref{CY34}) and such that the two form $\omega_4$ appearing in (\ref{sisv})
is of type $(1,1)$ with respect to the complex coordinates. The last assumption automatically implies that
the complex (3,0) form is given by (\ref{omga}). These, together with the Calabi-Yau condition,
determined completely the local form of the Calabi-Yau metric (\ref{kitai})-(\ref{kitai2}). The task is now to show that
the metrics \cite{Fayyazuddin}-\cite{Chinos} are under these hypothesis and therefore they are a particular case of
(\ref{kitai})-(\ref{kitai2}).

\subsection{Examples with isometries preserving the whole SU(3) strucuture}

      In this subsection the results of \cite{Fayyazuddin}-
\cite{osvaldo} are briefly reported, for more details about the proofs
we refer the reader to the original references. The solution generating techniques of \cite{Fayyazuddin}-
\cite{osvaldo} start with a hyperkahler structure $\widetilde{\omega}_i$ with $i=1,2,3$ and one of these
closed two forms, say $\widetilde{\omega}_i$ is deformed to a new
$y$-dependent two form
\be\lb{der}
\omega_4(y)=\widetilde{\omega}_1-d_4 d_4^c G,
\ee
while $\widetilde{\omega}_2$ and $\widetilde{\omega}_3$ are kept intact. This
2-form
plays the role of $\omega_4(y)$ in (\ref{sisv}). Here the operator $d^c=J_1\;d$ is constructed
in terms of the complex structure $J_1$ which is defined by $\widetilde{\omega}_1$ and the hyperkahler metric by the relation (\ref{expro}). In the expression (\ref{der}) $G$ denotes an unknown function
which varies on $M_4$ and which, in a generic situation, may depend also on the coordinate
$y$. If there is a Killing vector preserving the whole $SU(3)$ structure,
which corresponds to the case $K_1=0$ in (\ref{anzat2}),
then the SU(3) structure (\ref{sisv}) and (\ref{omga}) take the following form
$$
\omega_6=\widetilde{\omega}_1-d_4 d^c_4 G+dy\wedge (d\alpha+A),
$$
\be\lb{nova2}
\psi_{+}=H^{2}\widetilde{\omega}_3\wedge dy+\widetilde{\omega}_2\wedge (d\alpha+A),
\ee
$$
\psi_{-}=-H^{2}\widetilde{\omega}_2\wedge dy+\widetilde{\omega}_3\wedge (d\alpha+A),
$$
with $\Psi=\psi_-+i \psi_+$. Given the deformed structure (\ref{der}), the compatibility condition (\ref{condi})
imply that
\be\lb{dofini}
(\widetilde{\omega}_1-d_4 d^c_4 G)\wedge(\widetilde{\omega}_1-d_4 d^c_4 G)=H^2 \widetilde{\omega}_2\wedge \widetilde{\omega}_2,
\ee
and, as the wedge products appearing in the last equality are all proportional to the volume
form $dV(g_4)$
of the initial hyperkahler metric $g_4$,  the relation
\be\lb{defini}
(\widetilde{\omega}_1-d_4 d^c_4 G)\wedge(\widetilde{\omega}_1-d_4 d^c_4 G)=M(G)\;\widetilde{\omega}_1\wedge \widetilde{\omega}_1,
\ee
defines a non-linear expression $M(G)$ involving $G$. The CY condition (\ref{co}) applied to (\ref{nova2}) impose further constraints, which are explained in detail in \cite{osvaldo} and which we will
not reproduce here. The result is that the geometry is described in terms of a non-linear differential
equation determining the function $G$ and which involves the operator $M(G)$, these equation is \footnote{This equation strongly resembles the one found in \cite{Apostolov} for the G2 holonomy case.}
\be\lb{munja2}
G_{yy}=M(G).
\ee
In addition the explicit expression for the  SU(3) structure is completely determined in terms of $G$ as
$$
\omega_6=\widetilde{\omega}_1-d_4 d^c_4 G+dy\wedge (d\alpha-d^c_4 G_y ),
$$
\be\lb{nova3}
\psi_{+}=G_{yy}\;\widetilde{\omega}_3\wedge dy+\widetilde{\omega}_2\wedge (d\alpha-d^c_4 G_y ),\ee
$$
\psi_{-}=-G_{yy}\;\widetilde{\omega}_2\wedge dy+\widetilde{\omega}_3\wedge (d\alpha-d^c_4 G_y ).
$$
The generic form of the 6-dimensional Calabi-Yau metric
corresponding to this structure is given by
\be\lb{gonoro}
g_6=g_{4}(y)+G_{yy}\;dy^2+\frac{(d\alpha-d^c_4 G_y )^2}{G_{yy}},
\ee
where $g_{4}(y)$ is  the Kahler 4-dimensional metric corresponding to the
deformed Kahler structure $\omega_1(y)=\widetilde{\omega}_1-d_4 d^c_4 G$.

    It is important to remark that the metrics of this subsection are under the hypothesis
leading to (\ref{kitai})-(\ref{kitai2}). First of all, the two form $\omega_1(y)$ introduced in (\ref{der}) is of
type $(1,1)$ with respect to the complex coordinates which diagonalize $J_1$, this follows
from the fact that $\omega_1$ is of type (1,1) with respect to these coordinates, and the term $d_4 d_4^c G$ is also of this type.
The form $\widetilde{\omega}_2+i\;\widetilde{\omega}_3$
is kept intact and, for a closed hyperkahler structure, is of type $(2,0)$. Moreover
(\ref{der}) is  closed
with respect to $d_4$, which leads immediately to the condition (\ref{kahlericidad}).
In addition (\ref{nova2}) is of the
type (\ref{kitai2}). All this imply that the metrics (\ref{nova3})-(\ref{gonoro}) are a
subcase of the family of Calabi-Yau metrics described in section 2.3.

\subsection{The Fayyazuddin linearization}

   The family of SU(3) structures (\ref{nova3}) and
(\ref{gonoro}) found above are completely determined in terms of a single function $G$ which is a solution of (\ref{munja2}). This is a non-linear equation and the general solution
is not known, but it can be solved in some specific examples. The source of the non-linearity of the operator
$M(G)$ defined in (\ref{defini}) and (\ref{munja2}) is the quadratic term
\be\lb{bedford}
Q(G)=d_4 d_4^c G\wedge d_4 d_4^c G,
\ee
therefore the operator $M(G)$ will reduce to a linear one if $Q(G)$ vanish.  This will be
the case when the function $G$ is of the form $G=G(w, \overline{w})$ where $w=f(z_1, z_2)$
is an holomorphic
function of the coordinates $(z_1, z_2)$ which diagonalize the complex structure $J_1$
\cite{Bedford}. This affirmation may be justified as follows. By use of the simple expression
\be\lb{df}
d d^cG=2 \;i \;G_{i\overline{j}}\;dz_i\wedge d\overline{z}_j,
\ee
the quadratic term (\ref{bedford}) may be rewritten as
\be\lb{game}
Q(G)=-4\;(G_{1\overline{1}}G_{2\overline{2}}-G_{1\overline{2}}G_{2\overline{1}})\; dz_1\wedge d\overline{z}_1\wedge dz_2\wedge d\overline{z}_2.
\ee
But the functional dependence  $G=G(w, \overline{w})$ imply that
$$
G_{i\overline{j}}= w_i\; \overline{w}_{\overline{j}}\;G_{w\overline{w}},
$$
and by inserting this into (\ref{game}) gives $Q(G)=0$. This result may be paraphrased
as follows. If the function $G$ depends only on two complex coordinates $(w, \overline{w})$
then the quantity $d_4 d_4^c G$ is essentially a 2-form in two dimensions, therefore the
wedge product (\ref{bedford}) vanish identically.

      The situation described above is essentially the one considered by Fayyazuddin in the
reference \cite{Fayyazuddin} and, if suitable boundary conditions are imposed,
the resulting metrics give a dual description of D6 branes wrapping a complex
submanifold in a hyperkahler manifold.  A simple example is obtained when the initial
hyperkahler structure is the flat metric on $R^4$ and $G_{yy}$ varies over an arbitrary set of
2-dimensional hyperplanes inside $R^4$. There it was shown in \cite{Fayyazuddin}
that the resulting metrics are the direct sum of the flat metric in $R^2\simeq C$
and a general Gibbons-Hawking metric in dimension four \cite{Gibhaw}. These metrics
are of holonomy SU(2), which is a subgroup of SU(3). Our aim in the following
is to improve this situation and find Calabi-Yau metrics with holonomy \emph{exactly} SU(3) by use of this linearization.

\subsection{Calabi-Yau extensions of the 4-dimensional BKTY metrics}

   In the present subsection Fayyazuddin linearization explained above will be illustrated with an explicit example.
This linearization is performed in terms of an initial hyperkahler structure
and the one considered in references \cite{Rychenkova}-\cite{Bergshoeff} will be chosen by
simplicity, namely the distance element
\be\lb{gibando}
g_4=z\;dz^2+z\;(dx^2+du^2)+\frac{1}{z}(dt-x\;du)^2.
\ee
By denoting $V=z$ and $A=-xdy$ it is seen that (\ref{gibando}) takes the usual Gibbons-Hawking form  \cite{Gibhaw}, which means that it is hyperkahler
and with a tri-holomorphic Killing vector $K=\partial_t$. The solution (\ref{gibando})
corresponds to a superposition of
6-branes, which results in a linearly growing potential independent on the coordinates
$(x,u)$. In fact $V=z$ is the electric potential for a infinite plane with
constant density charge at $z=0$, for which the electric field is constant.

   The first difficulty for (\ref{gibando}) is that crossing the plane $z=0$ implies a
change in its signature. Something similar happens for instance, for the Taub-Nut metric
with negative mass parameter. The last one corresponds to a potential $V=1-1/r$ and has
a change of signature when crossing the region $r=1$. For the Taub-Nut metric the
explanation is that it is asymptotic to the Atiyah-Hitchin metric, which is complete
and regular. The change of the signature is an indication the the Taub-Nut approximation
breaks down for $r>1$. It is plausible to think that something similar happens for
(\ref{gibando}). In fact, there have been several approaches to interpret its meaning.
The authors of \cite{Bergshoeff} proposed to replace $z$ by $|z|$ in (\ref{gibando}).
They justify this procedure by interpreting the region $z=0$ as a source plane and the regions $z>0$ and
$z<0$ are the sides of a domain wall. The problem is that the metric in the surface $z=0$
is singular. Another idea was introduced in \cite{Rychenkova}. In that reference the authors
were able to identify an exact hyperkahler metric for which (\ref{gibando}) is the
asymptotic form. These authors observed that the coordinates $(x,u)$ may parameterize a
torus $T^2$ by  making the coordinate $t$ periodic such that the periods satisfy
$$
n=\frac{T_x T_u}{T_t},
$$
being $n$ an integer. The resulting manifold is a nilmanifold for which the curvature of
the connection pulled back to the $T^2$ satisfy the Dirac quantization condition
$$
\frac{1}{T_t}\int_{T^2}F=n.
$$
By defining the "proper time" $w=2 z^{3/2}/3$ one can write (\ref{gibando}) as
\be\lb{gibando2}
g_4=dw^2+(\frac{3w}{2})^{-\frac{2}{3}}(\sigma^3)^2+(\frac{3w}{2})^{\frac{2}{3}}((\sigma^1)^2+(\sigma^2)^2)
\ee
where $\sigma_k$ are left invariant forms on the Heisenberg group. The metric (\ref{gibando2}) for $n=2$ is in fact of the Gibbons-Hawking form, and it was conjectured in \cite{Rychenkova} that they describe the asymptotic form of some specific CY metrics found in \cite{Kobayashi}-\cite{Yau2} by Bando, Kobayashi, Tian and Yau (BTKY metrics). These metrics arise
as a degenerate limit of a K3 surfaces. The point is that K3 surfaces has 58 parameter
moduli space and as one
moves to the boundary of the moduli space the metric may decompactify while remaining
complete and non singular. The metric (\ref{gibando2}) is believed to describe the
asymptotic metric of a K3 surface in one of those limits of the parameters.

   Our task is now to extend (\ref{gibando}) to a CY six metric. We do not expect
the resulting metric to
be complete as the initial hyperkahler is just valid as an asymptotic expression. But this
example is illustrates clearly how the Fayyazuddin linearization applies in a generic case.
In order to use the linearization
a complex coordinate system for (\ref{gibando}) should be found.
A Kahler 2-form for this metric is
\be\lb{ahlop}
\omega=dt\wedge du-zdz\wedge dx,
\ee
and the corresponding complex structure has the following non zero components
$$
J_t^t=\frac{x}{z},\qquad J_t^u=\frac{1}{z},\qquad J_x^z=-1,
$$
\be\lb{ci}
 J_u^t=\frac{z^2+x^2}{z},\qquad J_u^u=\frac{x}{z},\qquad J_z^x=1.
\ee
A complex coordinate system $z_i$ with $i=1,2$ is then any choice for which the components
of the complex structure
take the form $\widetilde{J}^{\overline{i}}_{\overline{j}}=-\widetilde{J}^i_j=\delta_i^j$.
This amounts to find a coordinate change for which
$$
\frac{\partial x^a}{\partial z^i}J_a^b\frac{\partial z^j}{\partial x^b}=\delta_i^j,
$$
and the last equation is equivalent  following the following system
\be\lb{ser}
(J^{a}_b-i\delta_b^a)\partial_a z^i=0,\qquad i=1,2
\ee
It can be checked from (\ref{ci}) that the equations (\ref{ser}) are equivalent
 to the two following independent equations
$$
\partial_z z^i=-i \partial_x z^i,\qquad i\partial_u z^i=(z-ix)\partial_t z^i.
$$
Two independent solutions of the last system are given by $z^1=-x+iz$ and
$z^2=i\;u\;(z-ix)+t$.

        Now let us suppose that the function  $G$ in (\ref{munja2}) is of the form
$G=u^2+U(w, \overline{w}, u)$ and we choose $w=z_1$. Let us denote $U_{uu}=H^2$. If we further assume that $U$ does not depend on $x$ then by
taking the derivative of (\ref{munja2}) with respect to $u$ twice gives an equation for
$H^2$, namely
\be\lb{nosor}
\bigg(\frac{1}{z}\partial^2_z+\partial_{u}^2\bigg) H^{2}=0,
\ee
with solution
\be\lb{sulo}
H^2=1+\frac{m}{(4\;z^3+9\;u^2)^{\frac{1}{6}}}.
\ee
By integrating twice with respect to the variable $u$ and remembering
that $U_{uu}=H^2$ it follows that
\be\lb{sulon}
G=u^2-\frac{(\sqrt{2})^{5}}{15}\;m \;z^{\frac{5}{2}}\;\bigg[-1+\bigg(1+\frac{9\;u^2}{4\;z^3}\bigg)^{\frac{5}{6}}-\frac{15\;u^2}{4\;z^3} \; _{2}F_{1}[(\frac{1}{6}, \frac{1}{2}), (\frac{3}{2}), -\frac{9\;u^2}{4\;z^3} ]\;\bigg]
\ee
where $_{2}F_{1}$ denote a generalized hypergeometric function.
Now a simple calculation shows that $A=d_4^c G_u=-G_{uz}dx$ and this
together with (\ref{sulon}) gives
\be\lb{as}
A=-\frac{m\;u}{2\;z\;(9\;u^2+4\;z^3)^{\frac{1}{6}}}\;\bigg[\;-3+2^{\frac{2}{3}}\bigg(1+\frac{9\;u^2}{4\;z^3}\bigg)^{\frac{1}{6}}\;_{2}F_{1}[(\frac{1}{6}, \frac{1}{2}), (\frac{3}{2}), -\frac{9\;u^2}{4\;z^3} ]\;\bigg]\;dx.
\ee
Also a simple calculation shows that (\ref{der}) is in this case
\be\lb{pj}
\omega_1(u)=\omega_1-dd^c G=\omega_1-G_{1\overline{1}}\;dz_1\wedge d\overline{z}_1=\omega_1+\;G_{zz}\; dz_1\wedge d\overline{z}_1
\ee
$\omega_1$ given in (\ref{ahlop}) and in the last step we took into account that
 $z=i \overline{z}_1-i z_1$. The explicit expression of (\ref{pj}) is obtained from
 (\ref{sulon}), the result is
\be\lb{skl}
\omega_1(u)=\omega_1+\frac{m }{2 \;(4 + \frac{9 u^2}{
   z^3})^{\frac{5}{6}}\; z^2 \;(9 u^2 + 4 z^3)^{\frac{1}{6}}} \bigg[\;9\;2^{\frac{2}{3}} \; u^2 + 2 \; \bigg(\;2 \;2^{\frac{2}{3}} - \bigg(4 + \frac{9 u^2}{z^3}\bigg)^{\frac{5}{6}}\bigg) \;z^3\bigg]\;dz_1\wedge d\overline{z}_1.
\ee
The metric $g_4(u)$ in (\ref{gonoro}) is the one that correspond to the modified
Kahler potential (\ref{skl}) namely
$$
 g_4(u)=\frac{1}{z}(dt-xdu)^2+z\;(du^2+dz^2+dx^2)
 $$
  \be\lb{sev}
  +\frac{m }{2 \;(4 + \frac{9 u^2}{
   z^3})^{\frac{5}{6}}\; z^2 \;(9 u^2 + 4 z^3)^{\frac{1}{6}}} \bigg[\;9\;2^{\frac{2}{3}} \; u^2 + 2 \; \bigg(\;2 \;2^{\frac{2}{3}} - \bigg(4 + \frac{9 u^2}{z^3}\bigg)^{\frac{5}{6}}\bigg) \;z^3\bigg]\;(dz^2+dx^2)
    \ee
By collecting the results (\ref{sulo})-(\ref{sev}) it follows that the Calabi-Yau extension
(\ref{gonoro}) of the BTKY metric is
$$
g_6=\bigg(1+\frac{m}{(4\;z^3+9\;u^2)^{\frac{1}{6}}}\bigg)^{-1}\;(d\alpha+A)^2+\bigg(1+\frac{m}{(4\;z^3+9\;u^2)^{\frac{1}{6}}}\bigg) \; du^2+\frac{1}{z}(dt-xdu)^2+z\;du^2
$$
\be\lb{govinda}
 +\bigg\{\frac{m }{2 \;(4 + \frac{9 u^2}{
   z^3})^{\frac{5}{6}}\; z^2 \;(9 u^2 + 4 z^3)^{\frac{1}{6}}} \bigg[\;9\;2^{\frac{2}{3}} \; u^2 + 2 \; \bigg(\;2 \;2^{\frac{2}{3}} - \bigg(4 + \frac{9 u^2}{z^3}\bigg)^{\frac{5}{6}}\bigg) \;z^3\bigg]+z\bigg\}\;(dz^2+dx^2).
   \ee
with $A$ given in (\ref{as}). This example is a non-trivial Ricci-flat and Kahler
metric in six dimensions, with holonomy exactly SU(3). Nevertheless in the
region near $z=0$ we do not expect our solution to be valid, as the  approximation
(\ref{gibando}) breaks down.

\subsection{Complete examples with Hamiltonian isometries}

     The explicit examples presented in the previous sections do possess isometries
preserving the full SU(3) structure, in other words, they correspond to the case
$K_1=0$ of section 3.1. The remaining case $K_1=1$ was considered in \cite{Chinos}. These authors
propose an anzatz which is given  in terms of an initial hyperkahler structure which is deformed as in (\ref{der}). In addition
they propose a sympletic form $\omega_6$ of the form (\ref{nova2}). The unique difference with the case considered in section 3.1 is that the complex three form is now
$\alpha$-dependent and is given by $\Psi=e^{i\alpha}(\psi_-+i \psi_+)$, with $\psi_\pm$ given (\ref{nova2}).
This imply that the isometry preserves the Kahler 2-form but not $\Psi$. The compatibility and the Calabi-Yau conditions
were worked out explicitly in \cite{Chinos} and the outcome is again that the metric and the $SU(3)$ structure is completely determined by $G$, which is now a solution of the equation
\begin{equation}\lb{munja5}
(e^{-\frac{1}{2}G_y})_y=M(G),
\end{equation}
$M(G)$ being the non-linear operator defined by (\ref{defini}). The CY metric is again given by (\ref{gonoro}) but now G is a solution
of (\ref{munja5}). It has been shown in \cite{Chinos} that complete metrics may be obtained when the initial hyperkahler structure is the flat one. In this case (\ref{munja5}) becomes
\begin{equation}
(e^{-\frac{1}{2}G_y})_y
=2(1+G_{1\bar 1}+G_{2\bar 2}+G_{1\bar 1}G_{2\bar 2}-G_{1\bar
2}G_{2\bar 1}).\label{lrus}
\end{equation}
 By parameterizing
\begin{equation}\label{lrus2} z_1=r\;\cos\frac{\theta}{2}\;\exp(\frac{i(\psi+\phi)}{2}),\qquad
z_2=r\;\sin\frac{\theta}{2}\;\exp(\frac{i(\psi-\phi)}{2}),
\end{equation}
and assuming that $G$ is a function of $r$ and $y$ the equation (\ref{lrus2}) reduce to
\begin{equation}\label{lrusi}
(e^{-\frac{1}{2}G_y})_y
=\frac{1}{2\,r^3}\partial_r\left[r^4\left(1+\frac{1}{2r}\partial_r
G\right)^2\right], \end{equation}
which is the equation (61) of reference \cite{Chinos}.
Particular solutions of this equation has been found in that reference and which, after appropriate coordinate transformations and different rescalings, give the resulting family of Calabi-Yau metrics \cite{Chinos}
\be\lb{bunt}
g_6=\frac{dy^2}{W} + \frac{1}{4} W y^2 (d\alpha - s^2 \sigma_3)^2 +
y^2 \Big(\frac{ds^2}{V} + \frac{1}{4} V s^2 \sigma_3^2 + \frac{1}{4} s^2(
\sigma_1^2 + \sigma_2^2)\Big)
\ee
with
$$
W= 1 - \frac{a}{y^6}\qquad V=1 - s^2 - \frac{b}{s^4}.
$$
The metric with $b=0$ describes a higher
dimensional generalization of Eguchi-Hanson instanton \cite{Eguchi}-\cite{Gibbonsviejo}, with
$R^2\times CP^2$ topology and an asymptotic $R^6/Z_3$ \cite{popes}. For $a=0$,
the metric is a cone of $Y^{p,q}$. The general solution describes a
resolution of the $Y^{p,q}$ cone, and the detail global analysis can
be found in \cite{sun}-\cite{yi}. More details of this calculation
can be found in the original reference
\cite{Chinos}.

  It is important to remark that the equation (\ref{lrus}), which corresponds to the flat metric, is completely equivalent to (\ref{kitai}). This may be seen by making the redefinition
$G_{i\overline{j}}\to \delta_{i\overline{j}}+G_{i\overline{j}}$ in (\ref{kitai}), which gives (\ref{lrus}) as a result, and viceversa. In addition the complex coordinates $z_i$ appearing in (\ref{metricofin}) are locally given by (\ref{lrus2}). But although the starting point is the flat hyperkahler structure, it is not necessarily true that (\ref{lrus2}) parameterize $R^4$ globally, in fact there may appear singularities in the resulting Calabi-Yau metric which can be avoided by changing the periodicity of the angular variables or the range of the radial coordinate. In any case, the above reasoning shows that metrics (\ref{bunt}) are special solutions of  (\ref{kitai})
-(\ref{metricofin}).

    A priori, it may be expected that the use of curved hyperkahler backgrounds will enhance the number of solutions of (\ref{munja5}). In particular, it may sound plausible that if one starts with a gravitational instanton admitting a flat limit (such as the Taub-Nut one), then the resulting Calabi-Yau metrics obtained by solving (\ref{munja5}) will contain the ones arising from (\ref{lrus}) as a particular case and moreover, the families described by (\ref{lrus}) such as (\ref{bunt}) will be reobtained by taking the corresponding flat limit. As (\ref{lrus}) is equivalent to (\ref{kitai}) this reasoning will imply that (\ref{munja5}) describe a more general family that (\ref{kitai}). But what the results of the present work are showing is that the situation is the opposite, that is, any Calabi-Yau metric found in terms of a curved hyperkahler space by solving (\ref{munja5}) can be obtained from solutions of the "flat" equation (\ref{kitai}) as well. Thus the number of solutions of (\ref{munja5}) are less or equal to the solutions of (\ref{kitai}). The arguments showing this are the same than in the section 3.1 namely, that all the metrics described by (\ref{munja5}) are under the hypothesis giving the equation (\ref{kitai}). \footnote{See the last paragraph of section 3.1, in fact it is not difficult to see that the $\alpha$-dependent phase does not change these arguments at all.}  Although this conclusion may sound a bit odd, there is further evidence for that, which is the following. If one starts with a curved hyperkahler metric with tri-axial symmetry instead of the flat one, then Calabi-Yau metrics resulting from (\ref{munja5}) are the one with
$R^2\times CP^2$ topology and an asymptotic $R^6/Z_3$ together with the resolution of the cone over $T^{1,1}/Z_2$ \cite{Chinos}. But $T^{1,1}/Z_2$ is a particular case of the $Y^{p,q}$ Einstein-Sasaki manifolds thus, the solutions obtained with the tri-axial metrics are an special subcase of (\ref{bunt}). For other curved manifolds the system becomes harder to solve and no new solutions have been found. Although that formally there is nothing wrong with the use of hyperkahler structures to guess new solutions, it may be the case the use of curved geometries complicates the task instead of helping to solve it. For this reason it is perhaps convenient to find a formalism which avoid this problem, and the one developed here in (\ref{kitai})-(\ref{kitai2}) possess these advantages, as these equations do not make any reference to any vielbein of a curved hyperkahler metric.

\section{Discussion}

  In the present work, a family of Calabi-Yau manifolds with a local Hamiltonian Killing
vector, i.e , a Killing vector which preserve the metric together
with the Kahler form was characterized. It was assumed that the complex $(3,0)$-form
is of the form $e^{ik}\overline{\Psi}$, where $\overline{\Psi}$ is preserved by the
Killing vector as well, and that the space of the orbits of the Killing vector is,
for fixed value
of the momentum map coordinate, a complex manifold, in such a way that the complex
structure of the 2-fold is part of the complex structure of the 3-fold. Under these
assumptions, it was shown that the local form of the geometry is completely determined
in terms of a function $G$ satisfying the non-linear equation (\ref{kitai}) if the phase
$k$ is non-trivial or (\ref{kitai2}) if the phase $k$ is zero. It has been also pointed
out that the constructions given in \cite{Fayyazuddin}, \cite{osvaldo} and \cite{Chinos}
are included in this family.

   The advantages of this method are that, unlike the ones presented in
\cite{Fayyazuddin}, \cite{osvaldo} and \cite{Chinos}, it does not require a hyperkahler structure
as initial input. As it was discussed in section 2, it is only required
that the 4 dimensional manifold defined by the orbits of the Killing vector for fixed momentum map coordinate is a complex 2-fold, and the Calabi-Yau conditions imply automatically that it is Kahler. In fact, the equations (\ref{kitai}) and (\ref{kitai2}) for the function $G$ defining the six dimensional metric does not contains any reference to the vielbein of the complex 2-fold. In this form one may avoid the complications in the calculation of the local form of the geometry due to the non-trivial curvature of an initial hyperkahler geometry.

    It is perhaps better to compare this situation with known results in four dimensions.
Consider a 4-dimensional Calabi-Yau (hyperkahler) space, such that the Killing vector
preserve the Kahler form $\omega_4$ but not $\Omega_4$. As is well known, the general
local form of the Ricci-flat Kahler 4-metric is \cite{gegodas}-\cite{plebe}
\begin{equation}\label{gegodas} g_4=u_z
[e^{u}(dx^2+dy^2)+dz^2]+u_{z}^{-1}[dt+(u_{x}dy-u_{y}dx)]^2, \end{equation}
where $u$ is the solution of the equation
\begin{equation}\label{Toda} (e^u)_{zz} + u_{yy} + u_{xx}=0. \end{equation}
Equation (\ref{Toda}) is known as the continuum limit of the sl(n) Toda
equation and is called SU($\infty$) Toda equation. The three dimensional base metric,
namely
$$
g_3=e^{u}(dx^2+dy^2)+dz^2,
$$
is Einstein-Weyl \cite{Todd}-\cite{Ward}. But the general Einstein-Weyl equation is not
related to a Toda system, so these base metrics are Einstein-Weyl spaces of restricted type.
One may try to find solutions of (\ref{Toda}) by perturbing around a solution related to a
known Einstein-Weyl structure. This is not wrong, but optional. In the same way the
4-dimensional metric (\ref{metricofin}) is Kahler with Kahler potential $G$, but $G$
is of restricted type, given by solutions of (\ref{kitai}) or (\ref{kitai2}). One may
try to find a solution to these equations by perturbing around a known hyperkahler one,
as it was done in \cite{Fayyazuddin}-\cite{Chinos}, but this is optional as well.

   We also presented in section 3.3 an example which is obtained by means of the Fayyazuddin
linearization. This example has holonomy exactly SU(3), but it is not complete.
It may be interesting to see if it is possible to find complete metrics by means of this
linearization, which will correspond to D6 branes wrapping a complex 1-cycle inside a
hyperkahler. We hope to answer this question in the near future.
\\

{\bf Acknowledgement:} M. L and O.P.S are supported by CONICET (Argentina) and by the
ANPCyT grant PICT-2007-00849. When we were finishing the present note the work \cite{wano}
appeared, which probably has overlap with the results presented here.
\\

\end{document}